\documentstyle[twoside,11pt,epsfig]{article}

\catcode`\@=11
\long\def\@makefntext#1{
\protect\noindent \hbox to 3.2pt {\hskip-.9pt  
$^{{\tenrm\@thefnmark}}$\hfil}#1\hfill}			%CAN BE USED 

\def\thefootnote{\fnsymbol{footnote}}
\def\@makefnmark{\hbox to 0pt{$^{\@thefnmark}$\hss}}	%ORIGINAL 
	
\def\ps@myheadings{\let\@mkboth\@gobbletwo
\def\@oddhead{\hbox{}
\rightmark\hfil\tenrm\thepage}  
\def\@oddfoot{}\def\@evenhead{\tenrm\thepage\hfil
\leftmark\hbox{}}\def\@evenfoot{}
\def\sectionmark##1{}\def\subsectionmark##1{}}

\renewcommand{\thefootnote}{\fnsymbol{footnote}}

\newcounter{sectionc}\newcounter{subsectionc}\newcounter{subsubsectionc}
\renewcommand{\section}[1] {\vspace{25pt}\addtocounter{sectionc}{1} 
\setcounter{subsectionc}{0}\setcounter{subsubsectionc}{0}\noindent 
	{\twelvebf\thesectionc.\kern0.35cm #1}\par\vspace{8pt}}
\renewcommand{\subsection}[1] {\vspace{25pt}\addtocounter{subsectionc}{1} 
	\setcounter{subsubsectionc}{0}\noindent 
	{\twelvebf\thesectionc.\thesubsectionc\kern0.35cm #1}\par 
	\vspace{8pt}}
\renewcommand{\subsubsection}[1] {\vspace{25pt}\addtocounter{subsubsectionc}{1}
	\noindent
	{\twelverm\thesectionc.\thesubsectionc.\thesubsubsectionc\kern0.35cm 
	{\kern1pt\twelveit #1}}\par\vspace{8pt}}

\newcommand{\smalllineskip}{\baselineskip=11pt}

\def\ninecirc{
\begin{picture}(0,0)
\put(4.4,1.8){\circle{7.45}}
\end{picture}}
\def\ninecopyright{\ninecirc\kern2.75pt\hbox{\eightrm c}} 

\newcommand{\copyrightheading}[1]
	{\vspace*{-1cm}\baselineskip=11pt{\flushleft
	{\ninerm Fractals, #1}\\
	{\ninerm $\ninecopyright$\,\,\, World Scientific 
	 Publishing Company}\\
	 }}

\def\abstracts#1#2#3{{
	\centering{\begin{minipage}{5.0in}\tenrm\baselineskip=12pt
        \centerline{\twelvebf Abstract}
	\vspace{5pt}
	\parindent=0pc #1\par 
	\parindent=1pc #2\par
	\parindent=1pc #3
	\end{minipage}}\par}} 

\def\ARTICLES{\kern6.15cm\hbox{${\vcenter{\vbox{
	\hrule height 0.4pt width 4.562truein	    %TOP 
	\hbox{\vrule width 0.4pt 		    %MIDDLE
	height 0.6truein			    %HEIGHT
	\raise0.565cm\hbox{\kern1pc
	\seventeenbf Articles}}
	\hrule height 0.4pt width 4.562truein}}}$}}  %BOTTOM

\renewenvironment{thebibliography}[1]
	{\frenchspacing
	 \tenrm\baselineskip=12pt
	 \begin{list}{\arabic{enumi}.}
	{\usecounter{enumi}\setlength{\parsep}{0pt}
	 \setlength{\leftmargin 12.7pt}{\rightmargin 0pt} %FOR 1--9 ITEMS
	 \setlength{\itemsep}{0pt} \settowidth
	{\labelwidth}{#1.}\sloppy}}{\end{list}}

\newcounter{itemlistc}
\newcounter{romanlistc}
\newcounter{alphlistc}
\newcounter{arabiclistc}

\newcommand{\fcaption}[1]{
        \refstepcounter{figure}
        \setbox\@tempboxa = \hbox{\footnotesize{\bf Fig.~\thefigure\phantom{00}}#1}
        \ifdim \wd\@tempboxa > 6in
           {\begin{center}
        \parbox{6in}{\footnotesize\smalllineskip{\bf Fig.~\thefigure\phantom{00}}#1}
            \end{center}}
        \else
             {\begin{center}
             {\footnotesize{\bf Fig.~\thefigure\phantom{00}}#1}
              \end{center}}
        \fi}

\newcommand{\tcaption}[1]{
        \refstepcounter{table}
        \setbox\@tempboxa = \hbox{\footnotesize\bf Table~\thetable\phantom{00}#1}
        \ifdim \wd\@tempboxa > 6in
           {\begin{center}
        \parbox{6in}{\footnotesize\smalllineskip\bf Table~\thetable\phantom{00}#1}
            \end{center}}
        \else
             {\begin{center}
             {\footnotesize\bf Table~\thetable\phantom{00}#1}
              \end{center}}
        \fi}

\def\@citex[#1]#2{\if@filesw\immediate\write\@auxout
	{\string\citation{#2}}\fi
\def\@citea{}\@cite{\@for\@citeb:=#2\do
	{\@citea\def\@citea{,}\@ifundefined
	{b@\@citeb}{{\bf ?}\@warning
	{Citation `\@citeb' on page \thepage \space undefined}}
	{\csname b@\@citeb\endcsname}}}{#1}}

\newif\if@cghi
\def\cite{\@cghitrue\@ifnextchar [{\@tempswatrue
	\@citex}{\@tempswafalse\@citex[]}}
\def\citelow{\@cghifalse\@ifnextchar [{\@tempswatrue
	\@citex}{\@tempswafalse\@citex[]}}
\def\@cite#1#2{{$\null^{#1}$\if@tempswa\typeout
	{IJCGA warning: optional citation argument 
	ignored: `#2'} \fi}}

\def\fnt#1#2{\footnotetext{\kern-.3em
	{$^{\mbox{\sevenrm #1}}$}{#2}}}

\def\runninghead#1#2{\protect\pagestyle{myheadings}
\markboth{\protect\nineit\,\,\,\,\,#1\hfill}
{\hfill\protect\nineit #2\,\,\,\,\,}}
\font\seventeenbf=cmbx10      scaled\magstep3
\font\twelverm=cmr10      scaled\magstep1
\font\twelveit=cmti10     scaled\magstep1
\font\twelvebf=cmbx10     scaled\magstep1

\font\tenrm=cmr10
\font\tenit=cmti10
\font\tenbf=cmbx10

\font\ninerm=cmr9
\font\nineit=cmti9

\font\eightrm=cmr8

\font\sevenrm=cmr7

\catcode`@=11					%SIZE OF OPENING PAGE NO.
\def\ps@plain{\let\@mkboth\@gobbletwo
     \def\@oddhead{}\def\@oddfoot{\ninerm\hfil\thepage
     \hfil}\def\@evenhead{}\let\@evenfoot\@oddfoot}

\def\ps@myheadings{\let\@mkboth\@gobbletwo	%SIZE OF R/H NUMBER
\def\@oddhead{\hbox{}
\rightmark\hfil\ninerm\thepage}   
\def\@oddfoot{}\def\@evenhead{\ninerm\thepage\hfil
\leftmark\hbox{}}\def\@evenfoot{}
\def\sectionmark##1{}\def\subsectionmark##1{}}
\catcode`@=12

\begin{document}

\runninghead{Scaling in cosmic structures }
{Scaling in cosmic structures }

%----------------------------------------------------------------------------
\renewcommand{\thefootnote}{\fnsymbol{footnote}}      %USE SYMBOLIC FOOTNOTE

%----------------------------------------------------------------------------
\thispagestyle{plain}
\setcounter{page}{1}

\copyrightheading{Vol.~0, No.~0 (0000)}

\vspace{6pc}
\leftline{\phantom{\ARTICLES}\hfill}

\vspace{3pc}
\leftline{\hskip-0.1cm\vbox{\hrule width6.99truein height0.15cm}\hfill}

\vspace{2pc}
\centerline{\seventeenbf SCALING IN COSMIC STRUCTURES}
\baselineskip=20pt
\centerline{Luciano Pietronero, Maurizio Bottaccio, Marco Montuori}
\baselineskip=12.5pt
\centerline{\it Physics Department $\&$ INFM sezione di 
Roma 1,}
\centerline{\it Universit\`a "La Sapienza", 
P.le A. Moro 2, 00185 Roma, Italy. }
\vspace{0.08truein}
\centerline{Francesco Sylos Labini}
\centerline{\it D\'epartement de Physique Th\'eorique,
Universit\'e de Gen\`eve,}  
\centerline{\it 24, quai E. Ansermet, 1211 Gen\`eve  }
\vspace{0.36truein}
\abstracts{
Abstract  
The study of the properties of cosmic structures in the universe is one of the 
most fascinating subject of the modern cosmology research.
Far from being predicted, the large scale structure of the 
matter distribution is a very recent discovery, which continuosly 
exhibits new features and issues. 
We have faced such topic along two directions; 
from one side we 
have studied the correlation properties of the 
cosmic structures, that we have found 
substantially different from 
the commonly accepted ones.
>From the other side, we have studied the 
statistical properties of the very simplified system,
in the attempt to capture the essential ingredients 
of the formation of the 
observed strucures.
}{}{}

\vspace{0.78truein}
\baselineskip=14.5pt
\section{INTRODUCTION}
\noindent

The existence of a cosmic structures in the universe 
is one of the most fascinating findings of the two 
last decades in observational cosmology.
Galaxy distribution is far from homogeneous on small scale and     
large scale structures (filaments and walls) appear to be limited  
only by the boundary of the     
sample in which they are detected.     
There is currently an  
acute debate on the result of the statistical     
analysis of large scale features. 
In the past years we propose a new 
statistical approach, which has shown 
surprisingly a fractal structure,
extending from small scales to 

distance beyond  to $40 h^{-1} Mpc$ and 

with larger statistical uncertainity, 

even up to $100 h^{-1} Mpc$.
In the following we report our findings along two directions 
of investigations:
in the first part we describe the standard analysis and its 
limits of applicability. We describe our novel analysis and 
the results we got in the characterization 
of large scale structures.
Contrary to standard claims, this analysis is completely 
consistent with the results of the standard analysis: 
it's the interpreation of the latter which is radically different.
In the second part, we refer on a study of the 
dynamics of a very simple model of gravitational formation 
of structures.
We have analysed the evolution and the 
statistical spatial properties of a {\it N-body} system of point masses, 
interacting through gravity.
The system is arranged as to simulate an infinite system of particles and 
with very simple initial conditions.

\section{COSMIC STRUCTURES}
\noindent

The usual way to investigate the properties of the 
galaxy spatial clustering 
is to measure the two point autocorrelation function $\xi(r)$ 
\cite{tot69,pee73}.
This is the most used statistical tool, since it can be measured 
quite accurately with current redshift surveys.
$\xi(r)$ is the spatial average of the fluctuations in the galaxy 
number density at 
distance $r$, with respect to an homogeneous distribution 
with the same number of 
galaxies.

Consider a little volume $\delta V$ at position $\vec{r_{i}}$; let   
$n(\vec{r_{i}})$ be the density of galaxies in $\delta V$ and 
$<n>= N/V$ the density of galaxy in the whole sample.

The galaxy density fluctuations in $\delta V$ with respect to the 
average galaxy density $<n>$, i.e. the galaxy 
relative density fluctuations, is: 
\begin{equation} 
\frac{ n(\vec{r_{i}})-<n >}{<n>}=\frac{\delta n(r_i) }{<n>}
\end{equation}
The two point correlation function 
$\xi(r)$ \index{correlation function} 
at the scale $r$ is the spatial 
average of the product of the relative density fluctuations  
in two little volumes at distance $r$:
\begin{equation}
\label{f1}
\xi(r) = < \frac{\delta n(\vec{r_i}+r)}{<n>}\frac{\delta n(\vec{r_i})}{<n>}>_i
=\frac{<n(\vec{r_{i}})n(\vec{r_{i}}  + \vec{r})>_{i}}{<n>^{2}}-1
\end{equation}
where the average is performed over the sample. 
Roughly speaking, a set of points is correlated on scale $r$ if $\xi(r) > 0$; it is
uncorrelated over a certain scale $R$ if $\xi(r) = 0$ for $r>R$.
In the latter case the points are evenly distributed at scale $R>r$ or, in 
another words, 
they have an homogeneous distribution at scale $R>r$.
In the definition of $\xi(r)$, the use of the sample density 
$<n>$ as reference value for the fluctuations of galaxies is 
the {\it conceptual assumption} that the galaxy distribution is 
{\it homogeneous 
at the scale of the sample}.

Clearly such an approach is valid if 
the average density $<n>$ of the sample {\it is }
the average density of the distribution, or, in other words, 
if the distribution is homogeneous on the scale of the sample. 
For this reason, $\xi(r)$ analysis {\it assumes the 
homogeneity} and it is unreliable for {\it testing} it. 

In order to use $\xi(r)$ analysis, 
the density of galaxies in the 
sample must be a good estimation 
of the density of the whole distribution of galaxies.
This may either be true or not; 
in any case, it should be checked {\it before} 
applying $\xi(r)$ analysis \cite{cole92}. 

In addition to such criticisms, the usual interpretation of $\xi(r)$ 
measure is uncorrect for an another aspect; it is customary to define  
a characteristic scale for the correlations in any spatial 
distribution of points 
with respect the amplitude of the $\xi(r)$.
The {\it correlation length of the distribution} $r_0$ 
is indeed defined as the scale such that $\xi(r_0)=1$ \cite{pee93}.

Such a definition is uncorrect since, in statistical mechanics, 
the {\it correlation length} 
of the distribution 
is defined by how fast the 
correlations vanish as a function of the scale, i.e. by the 
functional form of the $\xi(r)$ and not by its amplitude .

The quantity $r_0$, then, does not concern the {\it scale} 
of fluctuations and it is not correct referring to this as a measure of the 
characteristic size of correlations \cite{gai99,gab00}.
According to the $\xi(r)$ definition, $r_0$ simply 
separates a regime of large fluctuations 
$\delta n/<n> \gg 1$ from a regime of small fluctuations 
$ \delta n/<n> \ll 1$: this is correct 
{\it if the average density of the sample $<n>$} {\bf \it is} 
{\it the average density of galaxy distribution }.

Such problems of $\xi(r)$ approach can be avoided analysing 
the spatial correlations of the data set without any a priori 
assumptions on the homogeneity scale of the data itself.
\cite {cole92} 

The way to perform such an unbiased analysis is to 
study the behaviour of the {\it conditional average number of galaxies} $<N(<r)>$ 
or the {\it average conditional galaxy density} $\Gamma^*(r)$ versus 
the scale $r$. The two quantities are respectively:

\begin{equation}    
\label{nr}    
<N(<r)> = B\cdot r^{D}    
\end{equation} 

where $N(<r)$ is the number of galaxies contained in a sphere of 
radius $r$ centered on a galaxy of the sample. 
$<N(r)>$ is the average of $N(<r)$  
computed in all the spheres centered on every 
galaxy of the sample.

\begin{equation}
\label{gamma}
\Gamma^*(r) = \frac{<N(<r)>}{4/3 \pi r^3 }=
\frac{3B}{4\pi}\cdot r^{D-3}
\end{equation}

$\Gamma^*(r)$ is then 
the corresponding average density of galaxies in spheres of radius $r$ 
\cite{cole92,syl98}:
The exponent $D$ is called {\it the fractal dimension} 
and characterises in a quantitative way 
how the system fills the space, 
while the prefactor $\:B$ depends 
on the lower cut-off of the distribution.

$<N(<r)>$ and $\Gamma^*(r)$ are the suitable statistical 
tools to detect the two-point correlation properties of 
a spatial distribution of objects and the possible crossover 
scale between different statistical distributions.

If the point distribution has a {\it crossover} 
to an homogeneity distribution at scale $R$, 
$\Gamma^*(r)$ shows a flattening toward a constant value at such scale.
In this case, the fractal dimension of Eq.(\ref{nr}) 
and Eq.(\ref{gamma}) has 
the same value of the dimension of embedding space $d$, $D=d$ 
(in three-dimensional space $D=3$)\cite{man83,cole92,syl98}. 

If this does not happen, the density sample will not correspond to 
the density of the distribution and it will show 
correlations up to the sample size.
The simplest distribution with such properties is a fractal
structure \cite{man83}.
A fractal consists of a system in which more     
and more structures appear at smaller and     
smaller scales and the structures at small     
scales are similar to the ones at large scales.    
The distribution is then self-similar. 
It has a value of $D$   
smaller than $d$, $D < d$.
In 3-dimensional space, $d=3$, a fractal has $ D <3 $ and 

$\Gamma^*(r)$ is a {\it power law}.
The value of $N(<r)$ largely fluctuates 
by both {\it changing the starting point}, from which we compute $N(<r)$, and 
{\it the scale $r$}.
This is due to the scale invariant feature of a fractal structure,
which does not have any {\it characteristic length} \cite{m97,man83}.
It is simple to show that if we analyse a fractal structure 
with $\xi(r)$, we can obtain a value for the 
{\it correlation length} $r_0$, which 
evidently does not have any relation with the 
correlation properties of the system.
In fact such a value is simply a fraction to the 
size of the sample under analysis. 
Larger is the sample size, larger is the corresponding $r_0$.

\begin{figure}[!tbp]
\centerline{
        \epsfxsize=12.0cm
        \epsfbox{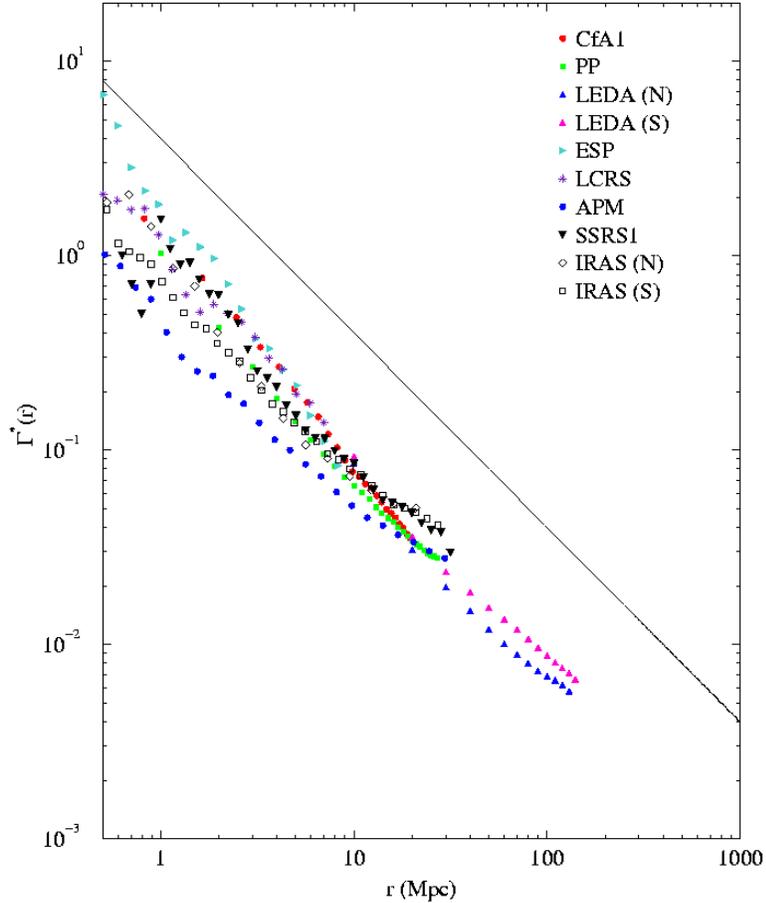}
        \vspace*{0.5cm}
        }
\caption{$\Gamma^*(r)$ in the range of scales  
$0.5 \div 100 h^{-1} Mpc$ for all the avalaible 3D galaxy data.
 A reference line with a slope   
$-1$ is also shown (i.e. fractal dimension $D = 2$).
}
\label{fig1}
\end{figure}

According to our criticism to the 
standard analysis, we have performed 
the measure of galaxy conditional average density 
$\Gamma^*(r)$ in all the   
three dimensional catalogs available. 
Our analysis is carried out on several 3D galaxy samples. 
The results are    
are collected in Fig. \ref{fig1}~\cite{syl98}. 

$\Gamma^*(r)$, measured in different   
catalogues, is a {\it power law} as a function of   
the scale $r$, extending from $\approx 0.5  $ to $30-40 h^{-1} Mpc $, 
without any tendency towards homogenization (flattening)\cite{syl98}.
In a single case, the LEDA sample \cite{diNella96}, 
it is possible to reach 
larger scales, $\sim 100 h^{-1} Mpc$.
The scaling $\Gamma^*(r)$ appears to continue with the same properties 
observed at smaller scales. 
This data sample has been largely criticised, but 
to our knowledge, never in a quantitative way. 
The statistical tests we performed show clearly that 
up to $50 h^{-1} Mpc$ the results are consistent 
with all other data \cite{syl98}.

Such results imply that the $\xi(r)$ analysis is 
inappropriate as it describes 
correlations as deviations from an assumed underlaying homogeneity.
The galaxy distribution shows instead fractal 
properties at least in the range $r \approx 0.5 - 30-40 h^{-1} Mpc $, which 
seem to extend in a single sample up $r \approx 100 h^{-1} Mpc $

By consequence, the {\it correlation length } $r_0$, i.e. the 
amplitude of the $\xi(r)$, for samples 
with such a linear extention, should be a fraction of the 
sample size:
$r_0$ should be larger for samples whose size is larger.

\begin{figure}[!tbp]
\centerline{
        \epsfxsize=12.0cm
        \epsfbox{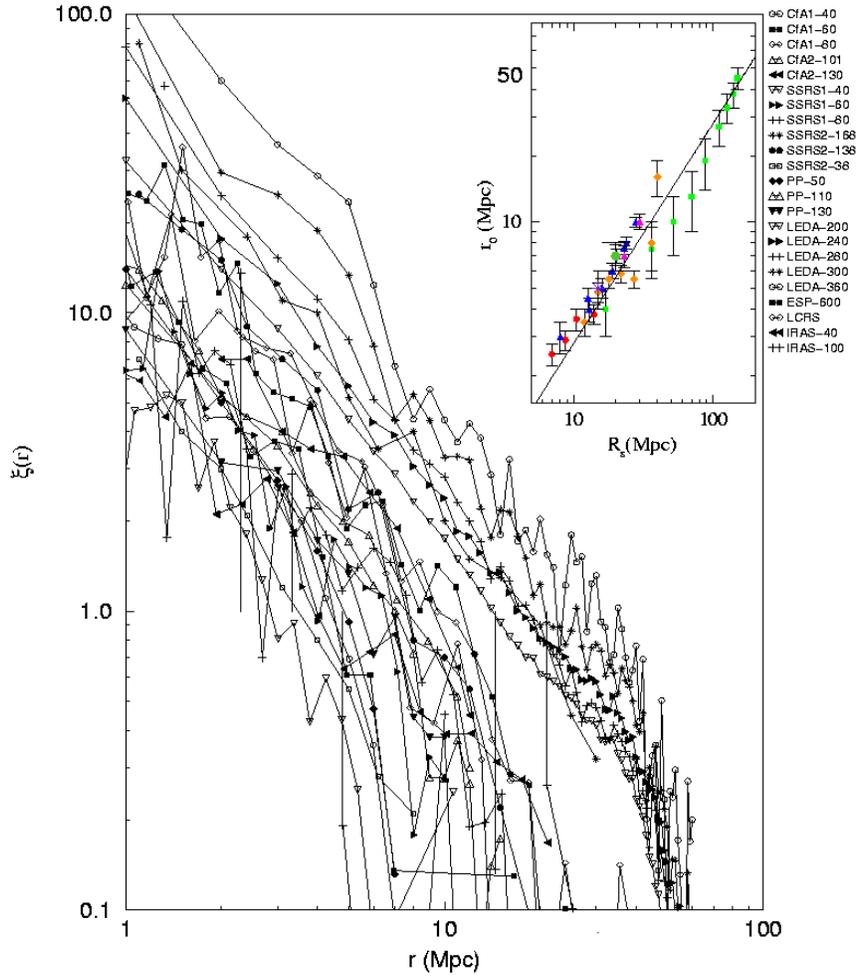}
        \vspace*{0.5cm}
        }
\caption{$\xi(r)$ measure in various VL galaxy samples. 
The general trend is an increase of the $\xi(r)$ amplitude 
for brighter and deeper samples.
In the {\it insert panel} we show the dependence   
of {\it correlation length} 
$r_0$ on {\it sample size} $R_s$ for all samples.  
The linear behaviour is a consequence   
of the fractal nature of galaxy distribution in these samples.
}
\label{fig2}
\end{figure}

This is evident in fig. (\ref{fig2}), where we plot the results 
of the standard analysis $\xi(r)$ performed on the same data sets 
analysed with $\Gamma^*(r)$ (fig. (\ref{fig1})).

In this case, then, $r_0$ has no relation with the correlation properties 
of the system; its variation in different samples is not related 
to any variation of the clustering of the 
corresponding data set.

\section{SIMULATIONS OF GRAVITATIONAL CLUSTERING}
\noindent
The study of the formation of the cosmic strucures 
we have analysed in the above section is one 
of the most challenging problems in astrophysics. 
Gravity is the most natural candidate for the 
explanation of the variety of structures we observe.
Indeed, the range of scales on which the gravitational 
clustering takes place is really 
impressive: from $10^{-1} pc$
to $10^8 pc$ ($1 pc = 3.2615 \, light-yr =3.0856 \, x \, 10^{18} \, cm$).
This implies interactions of gravity with other physical processes 
depending on the scale:  from turbulence in cold molecular clouds 
to cosmological expansion above galaxy cluster scale.
Because such a richness of physical processes can be involved 
in modelling the various structures we observe in cosmos, it is actually 
very difficult to retrieve a clear picture of the statistical properties 
of self-gravitating system.
Current astrophysical simulations have reached a high level of refinement, 
both in resolution 
and in the number of different physical processes which they take into account.
Such characteristics allow them to study in great detail the single physical 
problem for which they are developed  \cite{sim1}. 
On the other hand they don't allow a 
clarification the common role and the peculiarities of gravitational 
interaction.
On the contrary, we have 
tried to focus on such features analysing the most simplest 
case of {\it a many-body infinite self-gravitating system}, without 
any other ingredient but the gravity.
The theoretical approach to such a system 
goes back to Newton himself \cite{new62}, although it 
has faced by very few authors (e.g. \cite{saslaw}).
Indeed, the current theoretical effort is quite different 
since it is devoted to 
the study of evolution 
of a continuos gravitating fluid, which is assumed to have 
peculiar initial density fluctuations \cite{pee93,padma93}.  
>From the point of view of statistical mechanics, it is very
hard to study the properties of  an infinite system of self-gravitating 
particles. This is mainly due to the long range
nature of gravitational potential, which is not shielded
by the balance of far away charges, as e.g. in a  
plasma. Therefore all scales contribute to 
the potential energy of a particle. 
The peculiar  form of the gravitational potential 
produces two classes of problems: 
those due to 
the {\em short range} (i.e. $r\rightarrow0$)
divergence 
and those due to the {\em long range} 
(i.e. $r\rightarrow +{\infty}$) behaviour.
The former is not uncommon, since it  is the same problem
which arises in electromagnetism. The divergence would cause, e.g.,
the Boltzmann factor to diverge in the limit $r\rightarrow0$.
A typical prescription is to put a small distance 
cut-off in the potential. The physical nature of this cut-off may be 
due to many effects, e.g.  the dynamical
emergence of angular momentum barriers.
The long range behaviour is of much more concern and is, in fact,
{\em the} problem. 
It is an easy exercise to verify that the energy
of a particle in an infinite self-gravitating system diverges. 
This causes the energy to be {\em non-extensive}.
As a consequence,  a thermodynamical limit is not achieved,
since as the number of particles goes to infinity, even keeping 
the density constant, the energy {\em per particle} diverges.
Strangely enough, such a problem has not been fully appreciated
by  many physicists in the field (see e.g. \cite{Padma}), as they
try to avoid the long range divergence by putting the system in 
a box ``as it is usually done with ordinary gas''. In fact, the difference is
that in ordinary gas, when confining the system in a box, the energy
per particle is equal to a constant plus a surface term that goes
to zero in the thermodynamical limit. In self-gravitating systems,
due to non extensivity, the energy per particle is neither a constant,
nor the surface term goes to zero (in fact, it is of the same order of 
magnitude as the potential energy due to particles belonging to the system). 

Another very interesting consequence, which is often not appreciated,
is that the thermodynamical definition of temperature, as the parameter
which controls the equilibrium of the system, doesn't hold for a self 
gravitating system, since one cannot divide a system into smaller
subsystems with the same thermodynamic properties of the larger system. 

As a consequence of such difficulties, 
a satisfying thermodinamic equilibrium treatment
of such systems is still lacking.  

However we are much more interested in what happens {\it out of equilibrium},
during the evolution of a system.

The system we intend to simulate is a infinite many body system.
At this aim, the $N$ particles we effectively consider are 
confined in a cube of size $L$ submitted to periodic boundary conditions.
Every particle in the simulation box interacts with all other particles and 
with the periodic replicas of the whole system.

The initial conditions we consider are: \\
(1) {\it random (white noise) initial positions of particles};\\
(2) {\it no cosmological expansion};\\
(3) {\it zero initial velocities};\\
(4) {\it equal mass particles};\\

\begin{figure}[!tbp]
\centerline{
        \epsfxsize=12.0cm
        \epsfbox{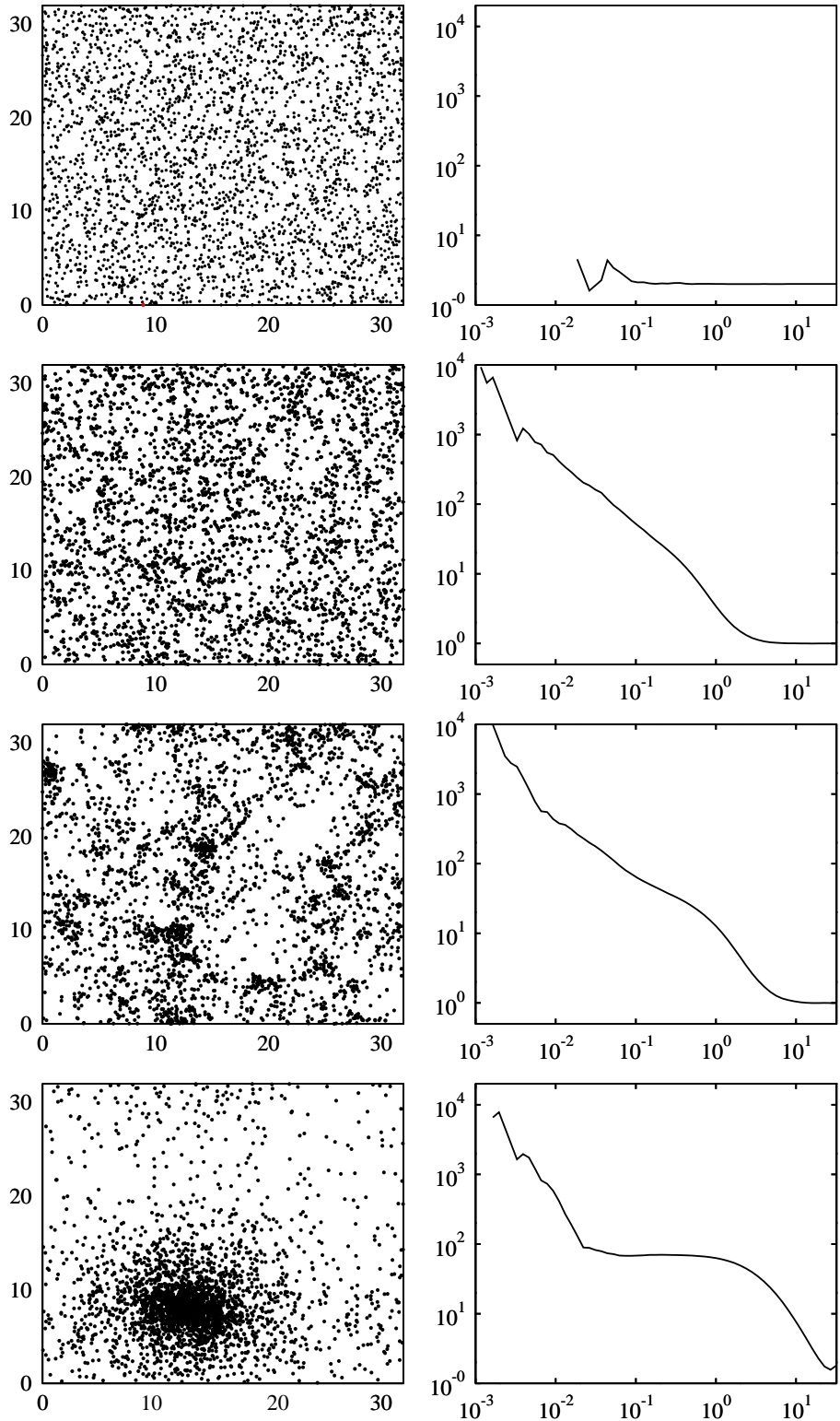}
        \vspace*{0.5cm}
        }
\caption{ {\it on the left}: snapshots of 
system evolution. The corresponding time is:
$t=0$, $t=2/3 \tau$, $t= \tau$ and $t= 4 \cdot \tau$
{\it on the right}: the corrisponding 
$\Gamma^*(r)$ for the  snapshots on the left
}
\label{fig3}
\end{figure}

Some snapshots of the temporal evolution of the system, 
with $N=32000$ particles, 
are shown in Fig. (\ref{fig3}).
The time evolution goes from the top to the bottom and 
the initial unclustered distribution of mass points 
evolves toward a clustered distribution. 
For each snapshot, we plot on the {\it right}, the corresponding 
$\Gamma^*(r)$. 
A typical time for the evolution of the system is 
$\tau = 1/\sqrt{G \rho}$, where $G$ is the gravitational constant and 
$\rho$ is the density of the system.
It is roughly the time needed to a particle to 
cross the system.
 
Fig. \ref{fig3} shows some interesting features, that we summarize 
in the following table.

\begin{table} [ht]
\begin{tabular}{|p{2.5cm}|p{6.5cm}|p{6.5cm}|} 
\hline
\bfseries $t$ time & Description & $\Gamma^*(r)$ \\
 
\hline 
\vspace{1cm}
$t_0=0$    & the system is composed by $N particles$ at 
rest with spatial possonian distribution in the simulation box  & 
the system has a 
costant number density $\Gamma^*(r,t_0)=1 $ at all the scales. 
At small scale, $\Gamma^*(r)$ is more fluctuating, because of  
the larger {\it poisson noise } $1/\sqrt{N}$ at small scale 
\\
\hline 
$t_1\approx 2/3 \tau $  &  the system starts to cluster at small scale &  
   $\Gamma^*(r,t_1)$ develops a larger amplitude at small scale     \\
 
\hline 
$t_2 \approx \tau$ & the clustering proces evolves with the merging 
of the small clusters in bigger ones &  the shape of $\Gamma^*(r,t_2)$ 
appears to be quite independent from time, but shifts 
toward larger scales for increasing time  \\
 
\hline  
$t_f \approx 4 \tau$  &  all the clusters have merged in a single big one, with 
dimension comparable with the simulation box size & 
$\Gamma^*(r,t_f)$ does not evolve anymore and the system has 
reached a stationary state \\
 
\hline 
\end{tabular} 
\caption{ {\it on the left}: projection of the simulation box 
onto $x-y$ plane at different time. {\it on the right}: corresponding 
measure of $\Gamma^*(r,t)$ at time $t$.}
\label{tabvlmock}
\end{table}

\section{CONCLUSIONS}
\noindent

The standard analysis of the correlation properties 
of the galaxy distribution is performed through the 
measure of the $\xi(r)$ function.
The latter can provide the correct information 
{\it if} the set under analysis is homogeneous inside the sample size.
For this reason $\xi(r)$ is not reliable for {\it testing} homogeneity.

This should be checked before the use of $\xi(r)$ and 
it is possible through $\Gamma^*(r)$ analysis of the sample set.
Such an analysis has been performed for the 
available 3D galaxy samples, with the result that the galaxy 
distribution appear fractal from $\approx 0.5$ to $100 h^{-1} Mpc$.

For such a range of scales, the  $\xi(r)$ analysis does not 
give the correct informations on the statistical properties of the 
galaxy distribution.

To investigate the formation of such a fractal structure in the 
universe, we are performing simulations of an N-body infinite 
self-gravitating system.
We started from the simplest initial conditions and 
analysed the system with the aforementioned $\Gamma^*(r)$ function.

Simulations with a different number of particles have shown 
that the shape of $\Gamma^*(r,t)$ for $t \le \tau$ 
is independent from $N$.
On the contrary, this is not true for 
the final state of equilibrium, when the system has formed a 
single cluster \cite{mb01}.

These measures seems to show that the transient phase, 

during which the collapse occurs, posses a well defined 

thermodynamical limit, which we are currently analysing \cite{mb01}.

The discrete nature of the N-body system seems to be a 

fundamental ingredient in the development 

of the spatial correlations.

The latter, indeed, grow at the small scale, where the discretness 

of the point distribution has to be taken into account.

At the moment is not clear if such a system can develop fractal 

correlations as seen in the galaxy distribution.

\section{ACKNOWLEDGMENTS}
\noindent
We would like to thank A.Gabrielli and M.Joyce for enlighting discussions.
We acknowledge financial support from EC TMR Research Network
under contract ERBFMRXCT960062 and the project 
{\it Clustering} by INFM.

\end{document}